\documentclass{article}

\usepackage{acronym}
\usepackage{authblk}
\usepackage{graphicx}
\usepackage{amsmath,amssymb,amsfonts}
\usepackage[title]{appendix}
\usepackage{subcaption}
\usepackage{cleveref}
\usepackage{nicefrac}
\usepackage{booktabs}
\usepackage{fullpage}
\usepackage{overpic}
\usepackage{bm}
\usepackage{xcolor}

\newcommand{\Rey}{\ensuremath{ \text{Re} }}

\DeclareMathSymbol{\shortminus}{\mathbin}{AMSa}{"39}

\crefname{figure}{Fig.}{Figs.}
\Crefname{figure}{Fig.}{Figs.}

\crefname{table}{Table}{Tables}
\Crefname{table}{Table}{Tables}

\crefname{equation}{Eq.}{Eqs.}
\Crefname{equation}{Eq.}{Eqs.}

\crefname{section}{Section}{Sections}
\Crefname{section}{Section}{Sections}

\begin{document}

\newacro{DNS}{direct numerical simulation}
\newacro{FDNS}{filtered direct numerical simulation}
\newacro{LES}{large-eddy simulation}
\newacro{SGS}{sub-grid scale}
\newacro{EKI}{ensemble Kalman inversion}
\newacro{SMARL}{scientific multi-agent reinforcement learning}
\newacro{DNN}{deep neural network}
\newacro{GCM}{global climate model}
\newacro{PDF}{probability density function}


\title{Prediction of Extreme Events in Multiscale Simulations of Geophysical Turbulence using Reinforcement Learning}
\author[1,2]{Yifei Guan} 
\author[3]{Lucas Amoudruz} 
\author[3]{Sergey Litvinov} 
\author[1,4]{Karan Jakhar} 
\author[1]{Rambod Mojgani} 
\author[3,$\dagger$]{Petros Koumoutsakos} 
\author[1,5,$\dagger$]{Pedram Hassanzadeh} 

\affil[1]{Department of Geophysical Sciences, University of Chicago, 5801 S Ellis Avenue, Chicago, 60637 IL, USA}
\affil[2]{Department of Mechanical Engineering, Union College, 807 Union Street, Schenectady, 12308 NY, USA}
\affil[3]{Computational Science and Engineering Laboratory, Harvard University, 29 Oxford Street Cambridge, 02138 MA, USA}
\affil[4]{Department of Mechanical Engineering, Rice University, 6100 Main Street, Houston, 77005 TX, USA}
\affil[5]{Committee on Computational and Applied Mathematics, University of Chicago, 5747 S Ellis Avenue, Chicago, 60637 IL, USA}
\affil[$\dagger$]{Corresponding authors: petros@seas.harvard.edu and pedramh@uchicago.edu}

\date{}

\maketitle

\begin{abstract}
Accurate subgrid-scale closures are essential for weather/climate models, where predicting extreme events is critical. Traditional closures have structural errors, e.g., producing excessive diffusion that dampens extremes. Artifical intelligence has gained attention for closure modeling, but the prediction of extreme events remains challenging. Supervised offline learning needs abundant high-fidelity training data and can lead to instabilities. Online learning algorithms are emerging as an alternative, but reliance on differentiable numerical solvers or scalable optimizers hinders broad use. Here, we introduce \ac{SMARL} to develop closures for canonical prototypes of atmospheric/oceanic turbulence, using only the enstrophy spectrum, estimated from a few high-fidelity samples,  as reward. This reward ensures that the model captures the cascades of scales in these simulations. These online-learned closures enable stable simulations, with up to five orders of magnitude fewer degrees of freedom, that reproduce high-fidelity simulation statistics and capturing in particularextremes. We interpret the  closures by analyzing the  \ac{SMARL} policy and  and demonstrate generalization to other flows. The results highlight  \ac{SMARL} as a potent tool for developing closures capable of capturing extremes in atmospheric/oceanic  flows, opening  new capabilities for effective climate modeling.

\end{abstract}

\section{Introduction}

Extreme weather events have enormous socioeconomic impacts~\cite{TOL2024,anchen_swiss_2025}, increasing the demand for the accurate estimates of their statistics, e.g., frequency, especially in a changing climate~\cite{ebi2021extreme, ipcc2021wgi, ipcc2022wgii}. However, due to their rarity and complex physics, high-fidelity simulations of these extreme events needs \acp{GCM} that are fast, thus of moderate to low resolution, while adequately representing the underlying multi-scale physics of the atmosphere, and more broadly, the Earth system. Practically fast-enough \acp{GCM}, which have typical resolutions of $\mathcal{O}(10)-\mathcal{O}(100)$~km, require \ac{SGS} closures (parameterizations) for many physical processes~\cite{randall2003breaking,hewitt2020resolving,fox2019challenges}. However, traditional physics-based closures have various parametric and structural errors, making them the main source of epistemic uncertainty in climate change predictions and weather forecasts, especially for extreme events~\cite{palmer2014climate,slingo2011uncertainty,lai2024machine,bracco2025machine}. Artificial intelligence (AI) applications for enhancing efficiency, e.g, through direct emulation of high-resolution simulations from \acp{GCM} or other models~\cite{ace2,bouabid2025score,li2024synthetic} or rare-event sampling with AI + \ac{GCM}~\cite{lancelin2025ai}, are emerging. However, these applications are in their early stages, and AI models have been shown to have challenges with the most extreme weather events, particularly those rarer and stronger than events in the training set, the so-called gray swans~\cite{sun_can_2025,lancelin2025ai}. Furthermore, even with these approaches, the need for high-fidelity closures for \acp{GCM} remains~\cite{Schneider_NCC_2023}.

Recently, there has been great interest in using AI, particularly \acp{DNN}, to develop better closures by learning directly from data. These AI-based approaches can be broadly categorized into two groups: supervised (offline) learning and online learning~\cite{bracco2025machine}. In supervised learning,  snapshots of high-fidelity data (e.g., from \ac{DNS}) are used as reference, then filtered and coarse-grained to extract the \ac{SGS} terms~\cite{grooms2021diffusion,guan2022stable, jakhar2023learning}.
The filtered data are then used to train a \ac{DNN} to learn the mapping between the filtered flow and the \ac{SGS} terms such as fluxes.
Once trained, the \ac{DNN} is coupled to a low-resolution numerical solver, the \ac{LES} hereafter, which represents a practical \ac{GCM}.
Studies with a variety of canonical prototypes of atmospheric and oceanic turbulence have shown the possibility of outperforming classical physics-based closures such as Smagorinsky and Leith models~\cite{maulik2019subgrid, bolton2019applications, guan2022stable,srinivasan2024turbulence, subel2022explaining}. In particular, it was found that these DNN-based closures not only reproduce the right amount of diffusion, but also capture backscattering, i.e., the forcing of the large-scale flow by the \ac{SGS} eddies, an important but challenging process for physics-based closures~\cite{Jansen_OM_2015, hewitt2020resolving}.

However, to work accurately, the supervised-learning approach has several major shortcomings: it is data hungry (a significant bottleneck for realistic systems), requires exact \ac{SGS} terms extracted from high-fidelity data (a non-trivial task), and can lead to unstable LES (with the root cause(s) remaining unclear)~\cite{o2018using, sun2023quantifying, bracco2025machine}. As a result, supervised-learning with DNNs faces significant challenges in scaling up beyond prototypes and to \acp{GCM}. More recently, supervised-learning with physics-constrained equation-discovery has shown promises as a data-efficient tool for developing accurate, generalizable, and interpretable closures~\cite{GM4}. However, the LES with these closures is not stable for very low resolutions including those that could be of practical interest in atmopshere and ocean modeling.


Online learning is emerging as a potent alternative to supervised learning, with the closure being learned by a DNN while it is  directly coupled to the low-resolution numerical (LES) solver.
Here, the  goal may not be to match spatially local flow quantities (such as the state's trajectory or the \ac{SGS} field), but rather, to match the low-order statistics of the high-fidelity simulations (e.g., DNS) or even (sparse)  observations.
In climate modeling, these statistics could be key properties of climate variability~\cite{pahlavan2023explainable} or kinetic energy spectra of turbulent flows~\cite{schneider2021learning, schneider2017earth}. Finally, online learning can account for the interactions between several closures, a common problem in climate modeling~\cite{lai2024machine} and unaddressed by supervised learning.
We distinguish three related approaches to online learning: 1) using a differentiable LES solver/\ac{GCM} that allows using backpropagation for optimization~\cite{frezat2022posteriori,Shankar_arxiv_2023,Gelbrecht_EGU_2022}, 2) using derivative-free ensemble-based optimization methods; e.g., \ac{EKI}~\cite{kovachki2019ensemble, dunbar2020calibration, schneider2020imposing, pahlavan2023explainable, mons2021ensemble, guan2024online}, and 3) using reinforcement learning~\cite{novati2021automating, Kurz_IJHFF_2023, Bae_NC_2022, kim2022deep}.
The first two approaches are actively pursued in the climate community with promising results, including the scale up of approach~(1) to realistic weather and climate prediction in Google's NeuralGCM~\cite{kochkov2024neural}. 
However, several major challenges remain.
For example, most current \acp{GCM} are not differentiable, and a differentiable solver requires a significant change in the modeling and software infrastructure~\cite{Schneider_NCC_2023}. 
Approach~(2) works with non-differentiable solvers, however, is largely limited to addressing parametric uncertainty but not structural uncertainty~\cite{guan2024online}.

\Ac{SMARL} has improved the parametric and structural uncertainties of closures for LES of 3D homogeneous and wall-bounded turbulent flows~\cite{novati2021automating,Kurz_IJHFF_2023,kim2022deep,Bae_NC_2022}.
However, its potential for climate modeling applications has remained unexplored. We note that the extension of models for homogeneous and wall-bounded flows to LES of geophysical turbulence and to \ac{GCM} is not straightforward. In LES of homogeneous turbulent flows~\cite{novati2021automating}, the states were based on specific well-known invariants of these flows. Similarly \ac{SMARL}-based wall models of turbulent flows required knowledge of the law of the wall to determine effective states and actions~\cite{Bae_NC_2022}.

\begin{figure}
 \centering
 \begin{overpic}
[width=\linewidth]{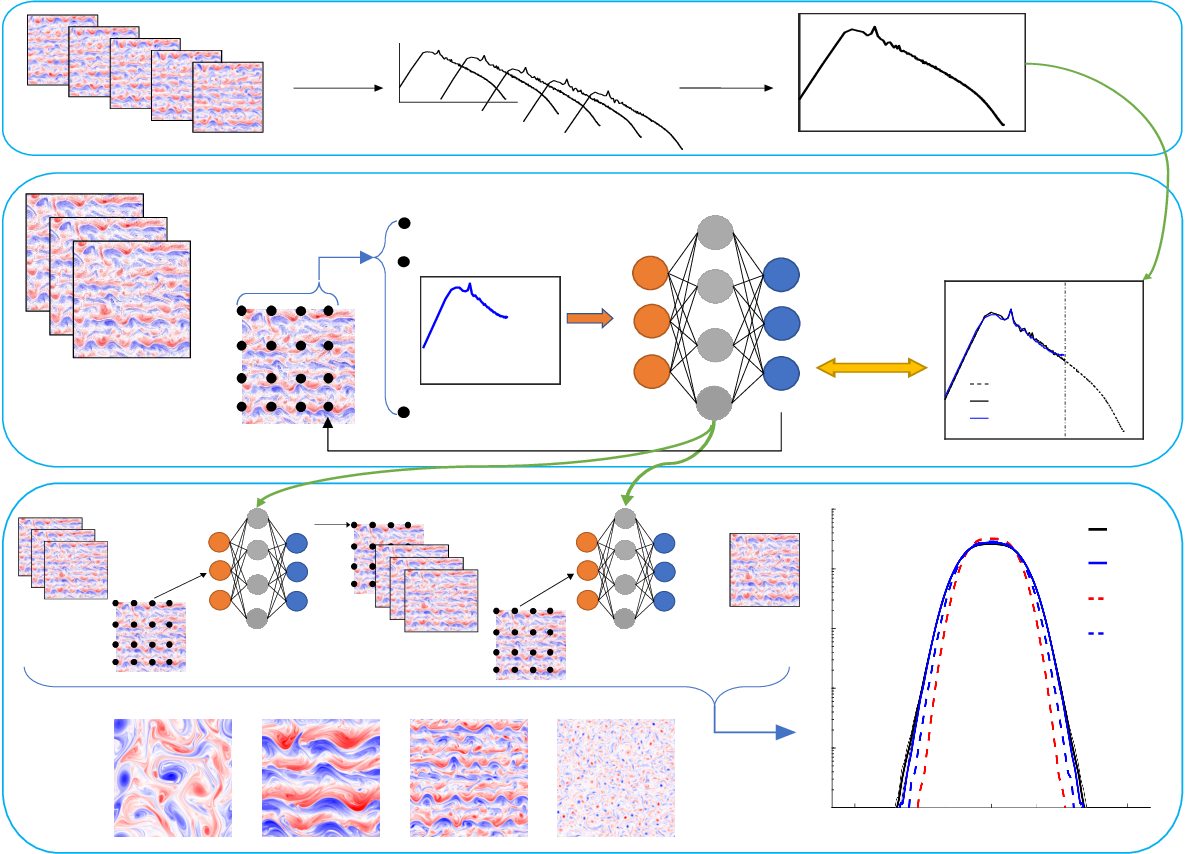}


\put(40,70) {\color{cyan} Reference Data}
\put(79,68.5) {Target}

\put(18,70) {\footnotesize $\omega$ snapshots $\times 5$}
\put(24,66) {\footnotesize $\hat{Z}=\frac{1}{2}\hat{\omega}^*\hat{\omega}$ }
\put(38,62) {\footnotesize $k$}
\put(59,66) {\footnotesize $\int_t$}

\put(62,68) {\footnotesize $\hat{Z}_\text{DNS}$}
\put(77,59.5) {\footnotesize $k$}

\put(1.5,64){\tiny $t_1$}
\put(5,63){\tiny $t_2$}
\put(8.5,62){\tiny $t_3$}
\put(12,61){\tiny $t_4$}
\put(15.5,60){\tiny $t_5$}


\put(18,55.5) {\small \color{cyan} Environment}
\put(15,53) {\small $\bar{\omega}$}
\put(1.0,45){\tiny $t_{j}$}
\put(2.7,43){\tiny $t_{j+1}$}
\put(5,41){\tiny $t_{j+2}$}
\put(15.5,40){\large$\cdot\cdot\cdot$}
\put(18.5,35.5){\tiny $t_{j+9}$}

\put(30,54.2) {\footnotesize Agents i}
\put(33.5,45) {\large$\cdot$}
\put(33.5,43) {\large$\cdot$}
\put(33.5,41) {\large$\cdot$}

\put(36.5,52.2) {\footnotesize Global state:}
\put(37,50) {\footnotesize $s'(t)=\hat{Z}_\text{LES}$}

\put(65,55.5) {\small \color{cyan} Training}

\put(69,44) {\tiny Low-resolution}
\put(69.5,42) {\footnotesize  LES solver}

\put(52.5,52.2) {\footnotesize States}

\put(63.5,52.2) {\footnotesize Actions}

\put(80,53) {\footnotesize Rewards: $r(t)=$}
\put(77,50.5) {\footnotesize $\frac{1}{||log(\hat{Z}_\text{FDNS})-log(\hat{Z}_\text{LES})||}$}

\put(84,39.3){\tiny DNS}
\put(84,37.9){\tiny FDNS}
\put(84,36.5){\tiny LES}

\put(85,33.5){\tiny $k$}
\put(90.5,37){\tiny $k_c$}

\put(30,34.7){\tiny $c_l(\bm{x}^{(i)},t)\sim\pi_w(\cdot|s'(t))$}

\put(10,25) {\small $\bar{\omega}$}
\put(0.8,22.2){\tiny $t_{0}$}
\put(2,21.2){\tiny $t_{1}$}
\put(3.2,20.2){\tiny $t_{2}$}
\put(5.5,18){\large$\cdot\cdot\cdot$}
\put(8.5,15){\tiny $t_{9}$}

\put(29,21.1){\tiny $t_{9}$}
\put(30.5,20.2){\tiny $t_{10}$}
\put(32,19.1){\tiny $t_{11}$}
\put(33.5,18.0){\tiny $t_{12}$}
\put(36,16){\large$\cdot\cdot\cdot$}
\put(39.5,15.2){\tiny $t_{19}$}

\put(57.5,23.5){\large$\cdot\cdot\cdot$}
\put(60.5,20){\tiny $t_{10^7}$}

\put(60.5,20){\tiny $t_{10^7}$}

\put(82,28){\footnotesize $\mathcal{P}(\omega)$}

\put(93.6,27){\tiny DNS}
\put(93.6,24.2){\tiny RL-Leith}
\put(93.6,21.2){\tiny DSmag}
\put(93.6,18.2){\tiny DLeith}

\put(74.5,2.5){\footnotesize $\shortminus4$}
\put(79,2.5){\footnotesize $\shortminus2$}
\put(83.2,2.5){\footnotesize $0$}
\put(87,2.5) {\footnotesize $2$}
\put(90.7,2.5) {\footnotesize $4$}
\put(82,1.0){\footnotesize $\omega/\sigma_\omega$}

\put(67.2,3.5){\tiny $10^{\shortminus6}$}
\put(67.2,8.5){\tiny $10^{\shortminus5}$}
\put(67.2,13.4){\tiny $10^{\shortminus4}$}
\put(67.2,18.5){\tiny $10^{\shortminus3}$}

\put(10,29){\color{cyan} Testing}
\put(3,6){Cases}

\put(14,11){\footnote{1}}
\put(27,11){\footnote{2}}
\put(39,11){\footnote{3}}
\put(51.5,11){\footnote{4}}

 \end{overpic}
 \caption{
  Training of SMARL for SGS closures: (1) reference data consists of 5 samples of a short DNS (\cref{eq:NS1,eq:NS2}); (2) online training of agents.
  The state-action map has input state $s'(t)$ (spectrum of enstrophy $\hat{Z}$ up to LES cutoff wavenumber $k_c$) and domain-averaged output action $c_l(t)$ (coefficient in the closure, \cref{eq:tau_SGS}) that maximizes the reward $r(t)$.
  During testing, the policy is coupled to the low-resolution numerical solver to produce a long LES that is $2000\times$ longer than the DNS training set and $1000\times$ the training horizon.
  Performance for extreme events is evaluated in terms of the vorticity PDF $\mathcal{P}(\omega)$, against DNS and dynamic physics-based SGS models.
 }
 \label{fig:overview}
\end{figure}
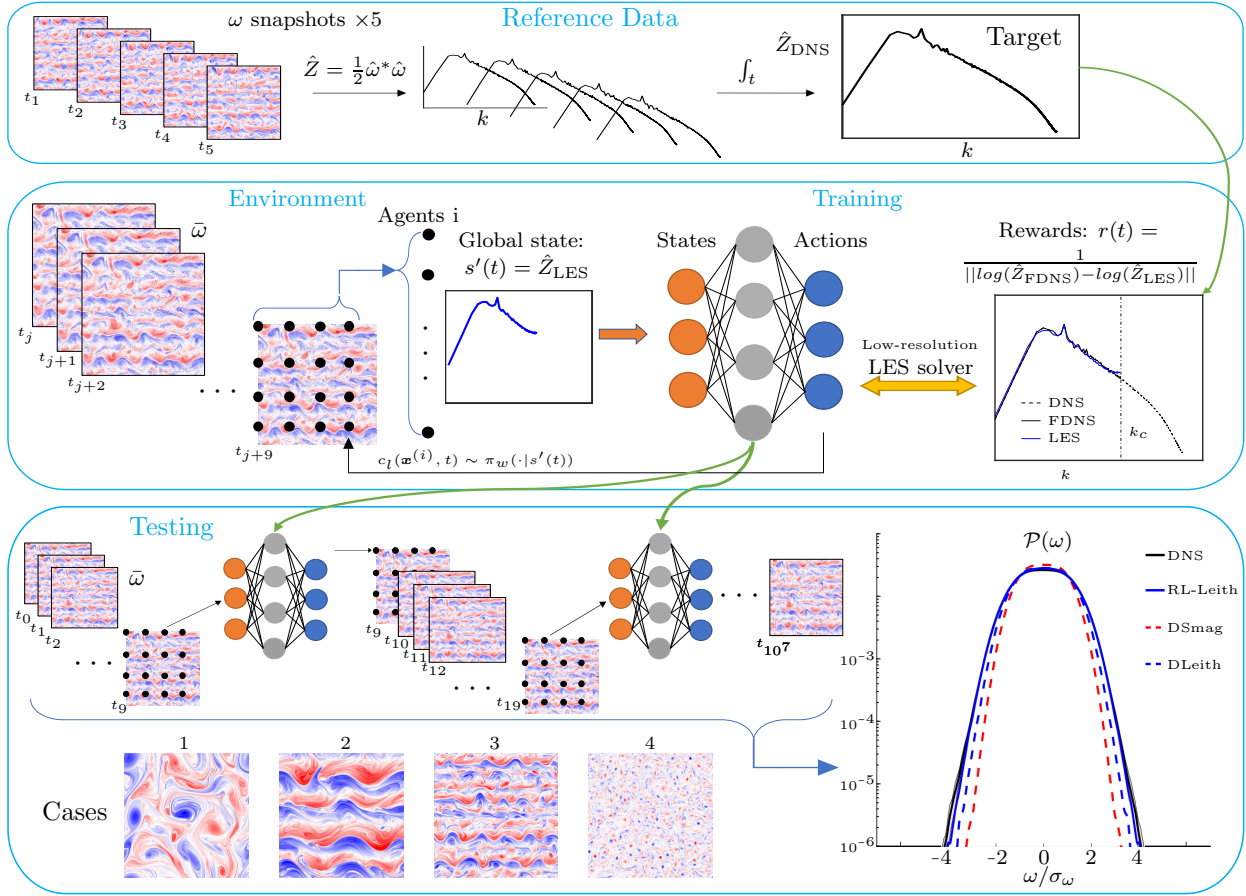

In this paper, we introduce the first application of reinforcement learning to closure modeling of geophysical turbulence. As summarized in \cref{fig:overview}, we apply \ac{SMARL} to a canonical prototype: several setups of 2D turbulence that contain a broad range of scales, structures (jets and vortices), and dynamics, well representing some of the key characteristics of atmospheric and oceanic flows (see Table 1 for details). 
Details of the approach can be found in Methods. Briefly,
\begin{enumerate}
\item [i] We train a \ac{SMARL}-based closure using low-order statistics. We  use  global states and targets to train a \ac{DNN} and learn the flow-dependent coefficient $c_l$ of a physics-based closure (Leith model) by matching the enstrophy spectrum of the \ac{LES}+\ac{SMARL} solver with that of a very short \ac{DNS}. These few training samples are far too few for DNN-based supervised learning. The LES solver does not need to be differentiable here.
\item [ii] We test the performance of this closure by comparing the kinetic energy spectrum and vorticity \ac{PDF} of the \ac{DNS} with  \ac{LES} having $160$ to $163,840\times$ spatio-temporally coarser resolution. The \ac{LES}+\ac{SMARL} spectrum and PDF match that of the DNS for simulations that are $\sim 2000\times$ the length of the training dataset. In particular, the PDFs match at the tails, which represent the rare extreme weather events in these prototypes. The \ac{LES}+\ac{SMARL} work at resolutions much lower than those required by the recently AI-discovered closure of~\cite{jakhar2025analytical}.
\item [iii] We analyze the interpretability of the \ac{SMARL}-based closures, using  Sobol indices to indicate the significance of contributions of the states to the actions. We also show that these closures learn both diffusion and backscattering, with the distribution of $c_l$ substantially differing from those estimated using dynamic Leith. This shows the ability of this approach in addressing both parametric and structural uncertainty.
\item [iv] We demonstrate successful generalization of the \ac{SMARL}-based closures on a  turbulent flow with $15\times$ higher $Re$ than its training \ac{DNS} data.
\end{enumerate}

Together, (i)-(iv) demonstrate the promises of \ac{SMARL} over supervised learning and other online learning approaches.


\section{Online learning of closures with SMARL}
Since the pioneering work of Smagorinsky~\cite{Smagorinsky_MWR_1963} for the closure modeling of geophysical turbulence, various eddy-viscosity models have been proposed. However, the Smagorinsky~\cite{Smagorinsky_MWR_1963} and  Leith~\cite{Leith_PhysicaD_1996} models remain today the most widely used closures. In these models, energy transfer from the resolved to subgrid scales is formulated  as non-linear diffusion~\cite{sagaut2006large}, often with dynamic coefficients that depend on space and time~\cite{germano1991dynamic,lilly1992proposed}.
The parameters of these models are estimated from local constraints, such as the Germano identity of energy transfer in dynamic Smagorinsky (DSmag), and the enstrophy transfer in dynamic Leith (DLeith).
The coefficients in Smagorinsky ($c_s$) and Leith ($c_l$) models are used to estimate the turbulence eddy viscosity: $\nu_e = c_s  \Delta^2 |\bar{\mathcal{S}}|$~\cite{smagorinsky1963general}, and $\nu_e  = c_l \Delta^3 |\nabla \bar{\omega}|$~\cite{leith1968diffusion,leith1996stochastic}, respectively.
Using these expressions, the turbulent shear stress, or subgrid stress, has the same form in both models~\cite{sagaut2006large,pope2001turbulent,meneveau2000scale}:
\begin{equation}\label{eq:tau_SGS}
  {\tau}^{\text{SGS}} = -2\nu_e \sqrt{2\mathcal{S}_{ij}\mathcal{S}_{ij}},  \;\mathcal{S}_{ij}=\frac{1}{2} \left( \frac{\partial u_i }{\partial x_j} + \frac{\partial u_j }{\partial x_i}  \right).
\end{equation}

 In this work, we use SMARL to learn a fully-connected \ac{DNN} that estimates the Leith coefficient ($c_l$) dynamically (also referred to as RL-Leith) and compare its performance to DLeith and DSmag (we have found RL-Leith and RL-Smag to perform similarly; see the appendix). In the context of the SMARL framework, the state of the system $s'(t)$ is defined by the enstrophy spectrum $\hat{Z}(k,t)$ up to the Nyquist wavenumber at LES (or cut-off wavenumber) $k_c$.
 The RL-Leith coefficients are thus expressed as $c_{l}(x,y,t) = f_\text{DNN}(\hat{Z}(k,t))$. 
 We place $n_{A,x} \times n_{A,y}$ agents distributed uniformly on the simulation grid.
 These agents produce actions that are interpolated onto the computation grid using a polynomial of $2^\textit{nd}$-order that preserves the periodicity of the domain. See Fig.~\ref{fig:overview} and Methods for the details of the SMARL setup, including the reward function (Eq.~(\ref{eq:reward})) and the training procedure.

\section{Results and Discussion}

We evaluated the proposed framework in five test cases, described in detail in \cref{se:cases}.
These benchmarks are widely used for atmospheric, oceanic, and planetary flows~\cite{boffetta2012two,perezhogin20202d,srinivasan2024turbulence,guan2022stable,Guan_PhysicaD_2023,jakhar2023learning}.
The simulations correspond to a two-dimensional flow with sinusoidal forced turbulence in a periodic domain.
This forcing term generates coherent structures of different scales.
In addition, some cases contain a Coriolis force $\beta$ that represents the rotation of the spherical Earth and generates jet streams.
The benchmarks are performed on a  range of Reynolds numbers covering different turbulence regimes.

\subsection{Prediction of extreme events}

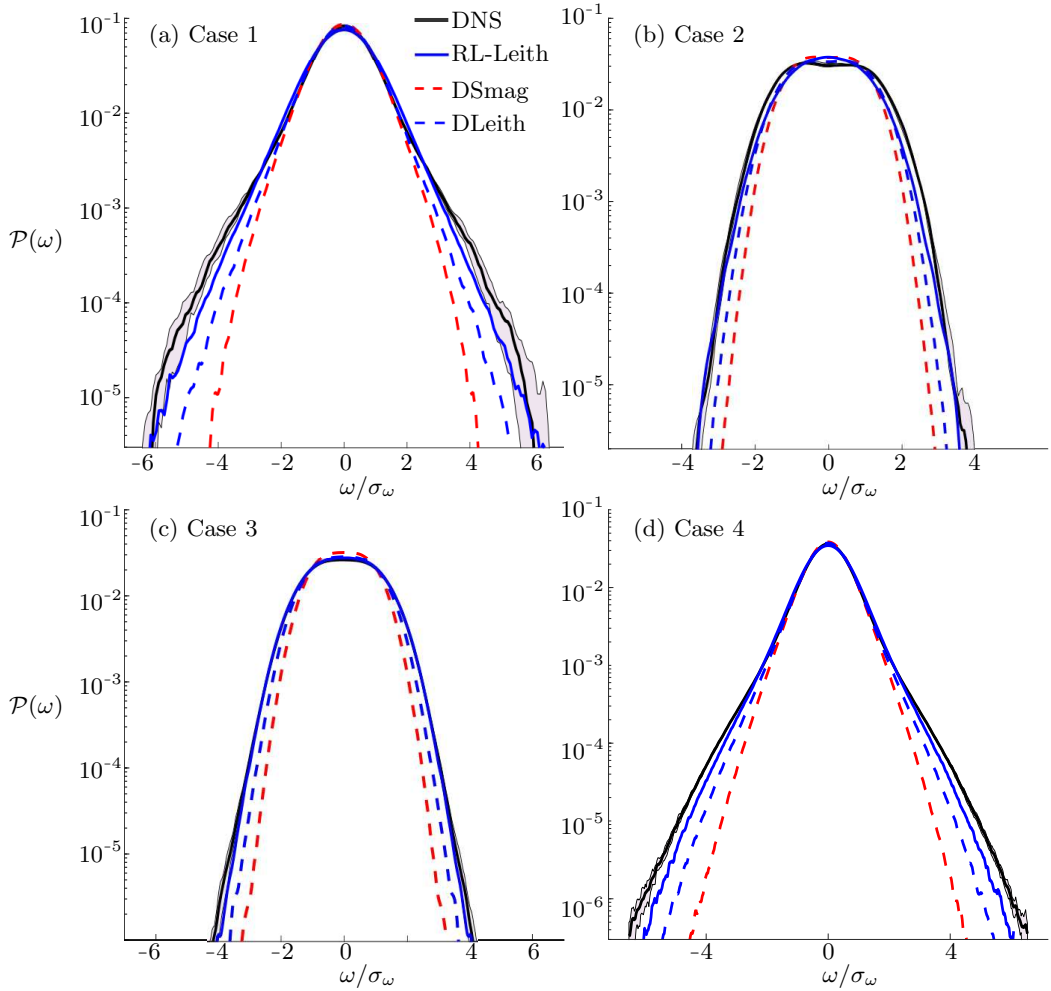
\begin{figure}
 \centering
 \begin{overpic}
[width=0.75\linewidth,height=0.75\linewidth]{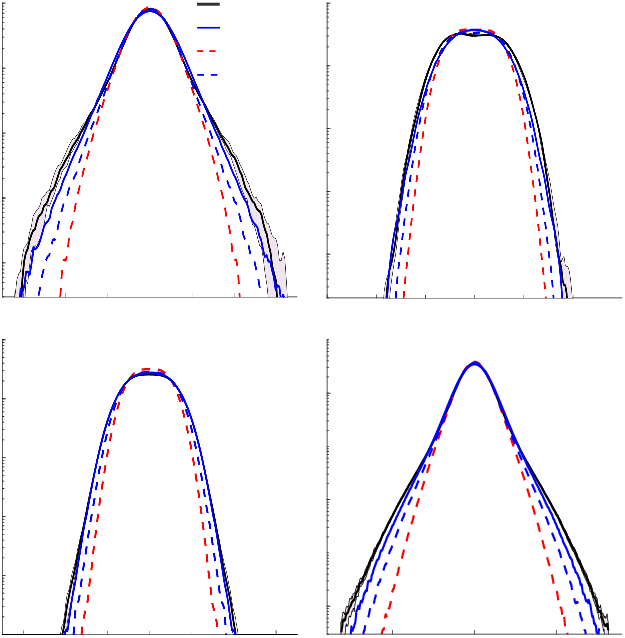}

\put(3,97){\small (a) Case 1}
\put(55,97){\small (b) Case 2}
\put(3,44){\small (c) Case 3}
\put(55,44){\small (d) Case 4}

\put(35.5,98.5){\small DNS}
\put(35.5,95){\small RL-Leith}
\put(35.5,91.0){\small DSmag}
\put(35.5,87.5){\small DLeith}

\put(1,51){\small $\shortminus 6$}
\put(9,51){\small $\shortminus 4$}
\put(16,51){\small $\shortminus 2$}
\put(23.5,51){\small $0$}
\put(30,51){\small $2$}
\put(37,51){\small $4$}
\put(44,51){\small $6$}

\put(-4.5,58){\small $10^{\shortminus 5}$}
\put(-4.5,68){\small $10^{\shortminus 4}$}
\put(-4.5,78){\small $10^{\shortminus 3}$}
\put(-4.5,88){\small $10^{\shortminus 2}$}
\put(-4.5,98){\small $10^{\shortminus 1}$}

\put(59,51){\small $\shortminus 4$}
\put(66.5,51){\small $\shortminus 2$}
\put(75.3,51){\small $0$}
\put(83,51){\small $2$}
\put(91,51){\small $4$}

\put(47.2,59){\small $10^{\shortminus 5}$}
\put(47.2,69){\small $10^{\shortminus 4}$}
\put(47.2,79){\small $10^{\shortminus 3}$}
\put(47.2,89){\small $10^{\shortminus 2}$}
\put(47.2,99){\small $10^{\shortminus 1}$}

\put(1.7,-1.5){\small $\shortminus 6$}
\put(9,-1.5){\small $\shortminus 4$}
\put(16,-1.5){\small $\shortminus 2$}
\put(23.5,-1.5){\small $0$}
\put(37,-1.5){\small $4$}
\put(43.7,-1.5){\small $6$}

\put(-4.5,9){\small $10^{\shortminus 5}$}
\put(-4.5,18){\small $10^{\shortminus 4}$}
\put(-4.5,27){\small $10^{\shortminus 3}$}
\put(-4.5,36){\small $10^{\shortminus 2}$}
\put(-4.5,45){\small $10^{\shortminus 1}$}

\put(61,-1.5){\small $\shortminus 4$}
\put(75.3,-1.5){\small $0$}
\put(88.5,-1.5){\small $4$}

\put(47.2,3){\small $10^{\shortminus 6}$}
\put(47.2,12){\small $10^{\shortminus 5}$}
\put(47.2,20){\small $10^{\shortminus 4}$}
\put(47.2,29){\small $10^{\shortminus 3}$}
\put(47.2,37){\small $10^{\shortminus 2}$}
\put(47.2,46){\small $10^{\shortminus 1}$}

\put(23.5,-4){\small $\omega/\sigma_\omega$}
\put(75.3,-4){\small $\omega/\sigma_\omega$}
\put(23.5,48.5){\small $\omega/\sigma_\omega$}
\put(75.3,48.5){\small $\omega/\sigma_\omega$}

\put(-12,25){\small $\mathcal{P}(\omega)$}
\put(-12,75){\small $\mathcal{P}(\omega)$}

\end{overpic}

\vspace*{5mm}

 \caption{
   {\it A-posteriori} testing results in terms of the PDF of vorticity ($\mathcal{P}(\omega)$). The $x-$axis is normalized by the standard deviation of vorticity ($\sigma_\omega$) computed from DNS of each case. The shading area represents the uncertainty (25-75 quartiles) of the DNS PDF. }
 \label{fig:aposteriori:testing}
\end{figure}

We quantify  extreme events by estimating the probability density function $\mathcal{P}(\omega)$ of the vorticity field $\omega$ over time and space. Positive (cyclonic) and negative (anti-cyclonic) $\omega$ in this prototype of atmospheric circulation correspond to low- and high-pressure weather systems, whose large amplitudes (tailes) drive various types of extreme events \cite{vallis2017atmospheric}.   \Cref{fig:aposteriori:testing} shows $\mathcal{P}(\omega)$ computed from cases 1-4 against the prediction performance of RL-Leith, DSmag, and DLeith.
In all cases, the RL-Leith model outperforms traditional physics-based dynamic models in terms of matching the DNS PDF.
Except for case 4, the RL-Leith PDF matches the DNS within uncertainties, even up at the tails of the distribution, corresponding to low probability or {\it extreme} events.
The lack of extreme events in the DSmag and DLeith predictions is attributed to the excessive diffusion from the positive clipping essential for stability \cite{guan2022stable}. This is consistent with  \cref{fig:apriori:testing} showing the excessive interscale enstrophy transfer predicted by DSmag and DLeith for all cases. Other {\it a-posteriori} analysis, including the turbulent kinetic energy spectra, is included in \cref{sec:aposteriori:kinetic:spectra}.

\subsection{Enstrophy transfer between scales}

\begin{figure}[tp!]
 \centering
 \begin{overpic}
[width=0.8\linewidth,height=0.4\linewidth]{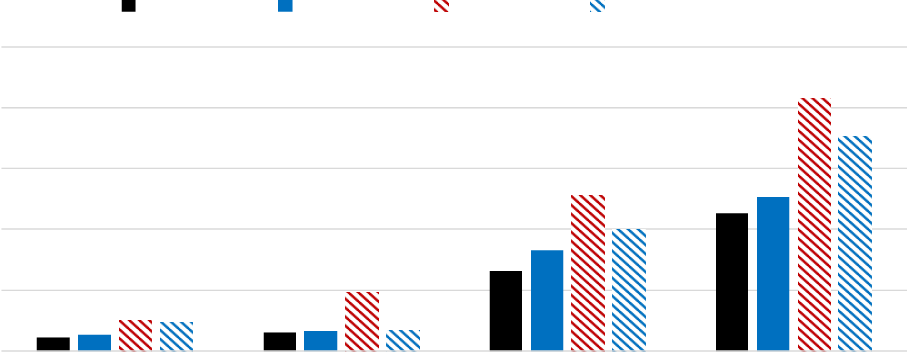}

\put(16.0,48.3){FDNS}
\put(33.0,48.3){RL-Leith}
\put(51,48.3){DSmag}
\put(68,48.3){DLeith}

\put(-3,-0.5){0}
\put(-3,8){5}
\put(-3,17){10}
\put(-3,25){15}
\put(-3,34){20}
\put(-3,42.5){25}

\put(-10,20){$\langle P_z\rangle$}

\put(8,-3){Case 1}
\put(34,-3){Case 2}
\put(58,-3){Case 3}
\put(86,-3){Case 4}

\end{overpic}
\vspace*{4mm}
 \caption{
   Total enstrophy diffusion $\langle P_z\rangle$ obtained from the FDNS, RL-Leith, DSmag and DLeith for the four cases.
   }
 \label{fig:apriori:testing}
\end{figure}

A main criterion for a functional eddy-viscosity LES model is the correct prediction  of the  the total energy/enstrophy transfer between resolved and subgrid scales~\cite{sagaut2006large}.
\Cref{fig:apriori:testing} shows the interscale enstrophy transfer ($\langle P_z\rangle = \langle \Pi\omega\rangle $) of RL-Leith, DSmag, and DLeith compared with the reference $\langle P_z\rangle$ obtained from filtering the \ac{DNS} results.
The RL-Leith model outperforms DSmag and DLeith in predicting the true enstrophy interscale transfer.
This is likely due to the ability of RL-Leith to capture backscattering (transfer from subgrid to resolved scales), resulting in less diffusion than DSmag and DLeith.
We also observe that DLeith outperforms DSmag in all 4 cases, demonstrating the benefits of using the Leith form of the eddy-viscosity model rather than the Smagorinsky form in LES of 2D turbulence, where capturing enstrophy interscale transfer is essential.
The advantage of DLeith over DSmag in terms of capturing the interscale enstrophy transfer is also reflected on {\it a-posteriori} analysis, see \cref{fig:aposteriori:testing}.

\subsection{Interpretability of the SMARL model}

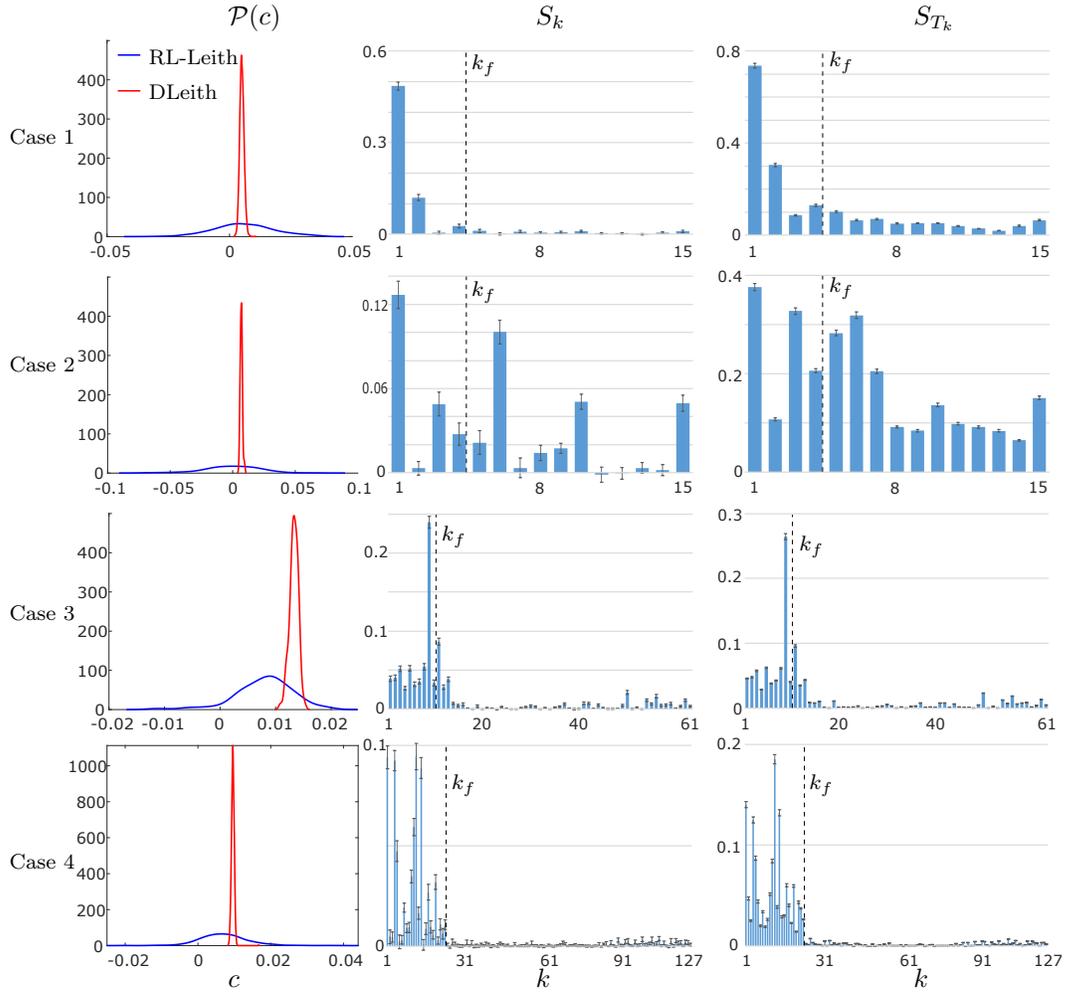
\begin{figure}[tp!]
 \centering
 \begin{overpic}
[width=0.8\linewidth,height=0.75\linewidth]{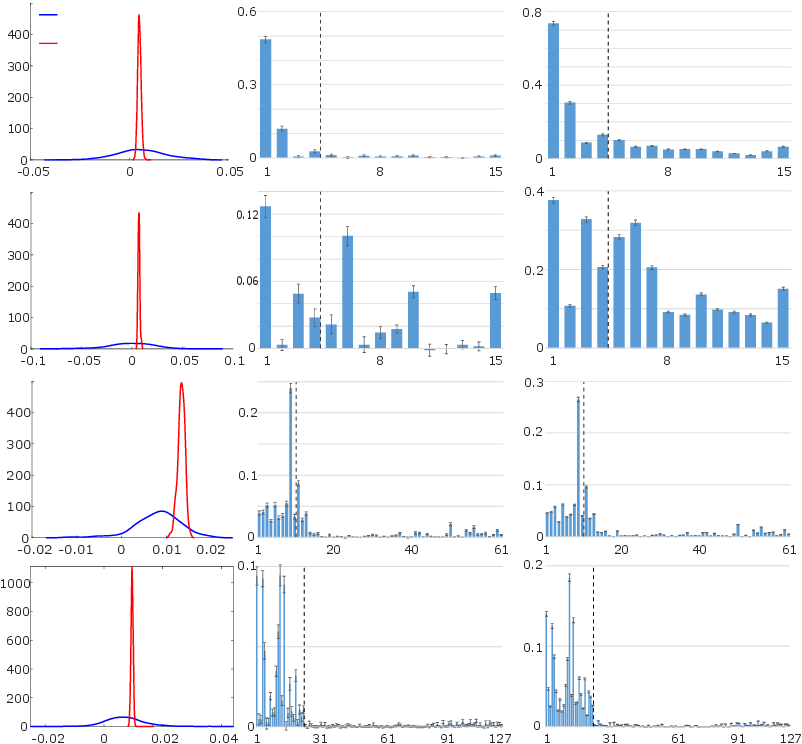}

\put(16.0,95){$\mathcal{P}(c)$}
\put(47.0,95){$S_k$}
\put(85,95){$S_{T_k}$}

\put(8,91){\footnotesize RL-Leith}
\put(8,87.5){\footnotesize DLeith}

\put(16,-2){$c$}
\put(47.0,-2){$k$}
\put(85,-2){$k$}

\put(-6,10){\footnotesize Case $4$}
\put(-6,35){\footnotesize Case $3$}
\put(-6,60){\footnotesize Case $2$}
\put(-6,83){\footnotesize Case $1$}

\put(40.5, 90.5){\footnotesize $k_f$}
\put(76.5, 90.8){\footnotesize $k_f$}
\put(40.5, 67.6){\footnotesize $k_f$}
\put(76.5, 68){\footnotesize $k_f$}
\put(37.5, 43){\footnotesize $k_f$}
\put(73.5, 42){\footnotesize $k_f$}
\put(38.5, 18){\footnotesize $k_f$}
\put(74.5, 18){\footnotesize $k_f$}

\end{overpic}
 \caption{
   First column: PDF of model coefficients $c$; second column: $S_k$; and third column: $S_{T_k}$. The error bars on the Sobol indices indicate 95\% confidence level. The Nyquist (or cut-off) wavenumber $k_c$ is the maximum wavenumber (i.e., $k_c=15$ in Case 1 and 2, $k_c=61$ in Case 3, and $k_c=127$ in Case 4). The dash lines in second and third columns mark the forcing wavenumber $k_f$.}
 \label{fig:state-action-map}
\end{figure}

To understand the superior performance of RL-Leith over the classical SGS models, we compare the distribution of Leith coefficients $c_l$ obtained from RL-Leith and DLeith.
The first column of \Cref{fig:state-action-map} shows the PDF of the Leith model coefficients $c_l$ for the four cases.
In all cases, the coefficients obtained with the DLeith model are concentrated around some positive values (rare negative values have been removed with positive clipping, which is essential for the LES stability with DLeith and DSmag \cite{maulik2019subgrid, guan2022stable}).
This leads to excessive diffusion, also confirmed by \cref{fig:aposteriori:testing,fig:apriori:testing}.
In contrast, the Leith coefficients learned from RL-Leith spread across a much wider range of values than DLeith, and include negative values (LES with Leith-RL remains stable despite these negative viscosities).
This shows that the RL-Leith model captures both the forward/diffusive and backscatter inter-scale energy/enstrophy transfer.
Similarly to DLeith, the majority of the coefficients predicted by RL-Leith is positive, showing that the RL-Leith model is more likely to be diffusive in general.
However, the highest probability density of the coefficients obtained with RL-Leith occurs closer to $0$, especially for Cases 3 and 4, indicating that the RL-Leith is less diffusive than DLeith.

The second and third columns of \cref{fig:state-action-map} show the sensitivity of the model to its inputs.
This is shown through the first-order and total Sobol indices of the model policy (see \cref{se:sobol} for details).
It decomposes the variance of the model output into contributions from: (1) each
individual input (first-order Sobol index $S_k$), and (2) interaction among inputs (total Sobol index $S_{T,k}$). In all cases, the model output is more sensitive to the low-wavenumber entries of the enstrophy spectrum.
This observation is consistent with the fact that most of the kinetic energy and enstrophy is contained in the large scales, corresponding to low wavenumbers $k$. For example, the ratio of turbulent kinetic energy $\langle \hat{E}(k<k_f)\rangle_k/\langle \hat{E}(k)\rangle_k=92.6\%$ for Case 1, $94.5\%$ for Case 2, $92.3\%$ for Case 3, and $93.4\%$ for Case 4.
In 2D turbulence, this region also coincides with the inverse energy cascade region, $k<k_f$, where $k_f$ is the forcing wavenumber of the 2D turbulence systems at which energy is injected into the flow (see~\cref{sec:methods}).
The next significant region is near the Nyquist wavenumber $k_c$, i.e. the maximum wavenumber in the LES simulations.
This region is responsible for interscale energy and enstrophy transfer between the resolved and unresolved scales.
Therefore, changes of $\hat{Z}$ at this wavenumber immediately trigger a change in the closure coefficient $c_l$. 
The Sobol indices suggest that the enstrophy contained in the middle wavenumber range does not have a significant impact on the leard coefficients $c_l$, and thus the learned viscosity.

\subsection{Generalization to higher Reynolds numbers}

\begin{figure}[tp!]
  \centering
 \begin{overpic}
[width=0.8\linewidth,height=0.4\linewidth]{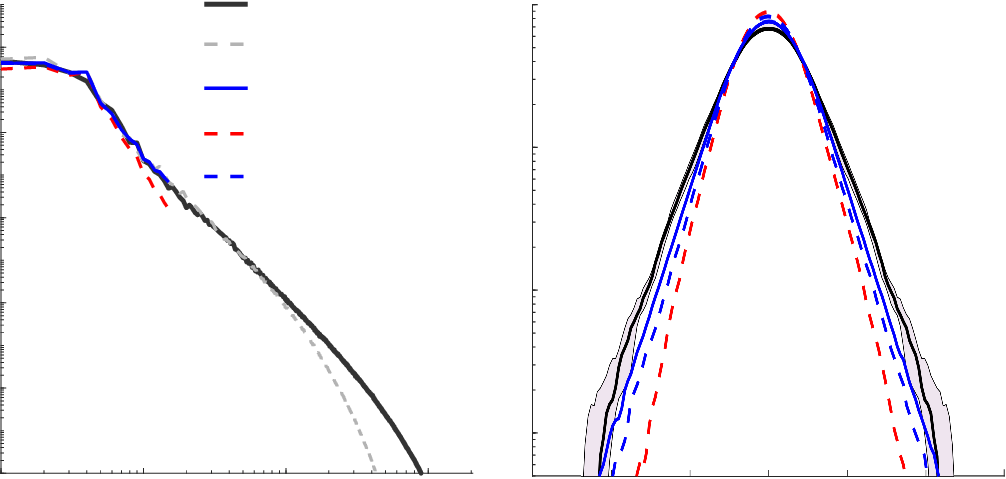}

\put(26,49){DNS (Case 5)}
\put(26,44.5){DNS (Case 1)}
\put(26,40){RL-Leith}
\put(26,35){DSmag}
\put(26,30.5){DLeith}

\put(2,48){(a) $\hat{E}(k)$}
\put(54,48){(b) $\mathcal{P}(\omega)$}

\put(20,-5){$k$}
\put(74,-5){$\omega/\sigma_\omega$}

\put(-4,44){$10^0$}
\put(-5,35){$10^{\shortminus 2}$}
\put(-5,26){$10^{\shortminus 4}$}
\put(-5,17){$10^{\shortminus 6}$}
\put(-5,9){$10^{\shortminus 8}$}
\put(-6,0){$10^{\shortminus 10}$}

\put(47.5,48){$10^{\shortminus 1}$}
\put(47.5,33.5){$10^{\shortminus 2}$}
\put(47.5,18.5){$10^{\shortminus 3}$}
\put(47.5,4){$10^{\shortminus 4}$}

\put(-1.5,-2){$10^0$}
\put(12,-2){$10^1$}
\put(26,-2){$10^2$}
\put(41,-2){$10^3$}

\put(59.2,-2){$\shortminus 4$}
\put(67,-2){$\shortminus 2$}
\put(75.7,-2){$0$}
\put(83.5,-2){$2$}
\put(91.4,-2){$4$}

\end{overpic}
\vspace{0.5cm}
 \caption{
   Generalization testing of SMARL to higher ($15\times Re$) turbulence (SMARL trained on Case 1 and tested on Case 5). (a) shows the turbulent kinetic energy spectra, $\hat{E}(k)$, and (b) shows the PDF, $\mathcal{P}(\omega)$. In (b), the $x-$axis is normalized by the standard deviation of vorticity ($\sigma_\omega$) computed from DNS. The shading area represents the uncertainty (25-75 quartiles) of the DNS PDF.
 }
 \label{fig:generalization}
\end{figure}
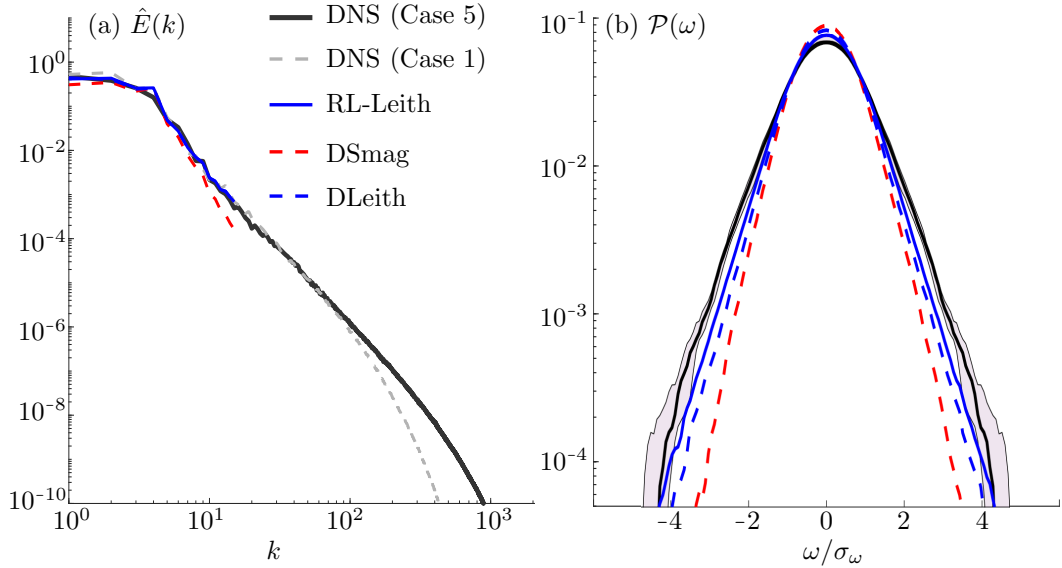

The generalization capability of  data-driven SGS models across different flow regimes (e.g., higher $Re$ turbulent flows) is essential for their practical use. However, many deep-learning-based SGS models do not generalize well and often require extra data or training efforts, such as transfer learning~\cite{guan2022stable,subel2022explaining} to be successful. Training data for high-$Re$ turbulent flows are especially difficult to obtain. For example, Case $5$ has $15\times$ higher $Re$ than Case $1$, and therefore \ac{DNS} to resolve Case $5$ requires $16\times$ higher spatial resolution compared to Case $1$ and correspondingly higher temporal resolution due to Courant-Friedrichs-Lewy condition. In \Cref{fig:generalization} we show that
the RL-Leith closure trained on Case $1$ can be directly used for Case $5$ without any new data from Case $5$ or an additional  training process.
It can be observed that the RL-Leith trained on Case $1$ is fully generalizable to work on LES of Case $5$. Similar to the testing for Case $1$, the RL-Leith model outperforms DSmag in terms of matching the \ac{DNS} kinetic energy spectra, $\hat{E}(k)$, and it outperforms both physics-based models (DSmag and DLeith) in matching the \ac{DNS} $\mathcal{P}(\omega)$ for Case 5 down to the tails (extreme events). The generalization capability of \ac{SMARL}-based SGS models to higher $Re$ turbulence can be explained by the state-action map that is trained during the RL online learning. As the state, $s'(t)$, is the enstrophy spectrum, $\hat{Z}(k,t)$, the difference between $Re=20,000$ (Case 1) and $Re=300,000$ (Case 5) lies only in the smallest scales that can only be resolved in \ac{DNS}. When the LES uses an aggressive (small) cut-off wavenumber (e.g., here for both Case 1 and 5, $k_c=16$), the cut-off $\hat{Z}_\text{DNS}(k,t)$ is very similar between Case 1 and 5, see \cref{fig:generalization} for the comparison between the energy spectrum, $\hat{E}(k)$, of Case 1 and 5.


\section{Conclusions}
\label{sec:conclusion}

We have introduced a data-efficient, \ac{SMARL}-based framework to learn LES closures for geophysical turbulence with broad potential applications to climate modeling. 
This framework is superior to existing AI-based closure modeling approaches as it works with: (i) sparse amounts of training data that are insufficient for supervised learning based on DNN~\cite{guan2023learning} and (ii) low LES resolutions inaccessible to closures identified with equation discovery~\cite{jakhar2023learning}. 
Furthermore, the framework  does not need a differentiable numerical solver and implicitly addresses model structural errors~\cite{guan2024online,heimbach2026reinforcement}.

We show that these closures enable LES to produce statistics, including energy spectra, PDF, and most importantly, tails of the PDFs, that closely match those of the far more expensive DNS.
Moreover,  LES with SMARL-based closures significantly outperform LES with classical physics-based closures in capturing extreme events, which are often diffusive to ensure stability but then underestimate extremes.
With a small number of samples from DNS, \ac{SMARL} develops closures capable of capturing the statistics of such extreme events, suggesting that both diffusion and backscattering, i.e., the two-way interaction of the resolved scales and SGS, are accurately represented. We emphasize that the \ac{LES} cases investigated here have $160$ to $163,840\times$ spatio-temporally coarser resolution compared to their DNS. LES+\ac{SMARL} is stable and can easily run thoughts of times longer than the lengths of the training set and training horizon.   

We provide an interpretation of the \ac{SMARL}-based closure by analysing the state-action map learned using \ac{SMARL} with the Sobol indices. Their distribution  across wavenumber, $k$ in the  enstrophy spectrum indicate that large-scale structures that contain most of the kinetic energy and enstrophy, have the most contribution to the prediction of the closure's coefficient $c_l$. The next significant contribution stems from the high-wavenumber region near the cut-off wavenumber, $k_c$. This can be interpreted as the directly responsible region of interscale energy and enstrophy transfer.

We have tested the generalization of the \ac{SMARL}-based closure by testing a pre-trained \ac{DNN} to a different flow regime where the $Re$ is increased by $15\times$. LES test shows that the \ac{SMARL}-based closure, albeit trained at a lower $Re$-number flow, performs better than DLeith and DSmag in terms of matching the kinetic energy spectrum and the PDF of vorticity, $\mathcal{P}(\omega)$, even at the tails of the distribution, corresponding to extreme vorticity events.

In summary, we demonstrate a \ac{SMARL}-based framework  to develop novel, accurate, stable, and generalizable SGS closure for geophysical turbulence, which can be scaled up to GCMs and Earth system models in the future.

\subsection*{Acknowledgments}

We thank Pascal Weber (Manukai) for the discussions on Korali.
PK and PH gratefully acknowledge support from the Salata Institute Climate project (GSD)-032549 and NSF CSSI Program (OAC-2005123), respectively.

\clearpage
\section{Methods}
\label{sec:methods}

\subsection{Direct and large-eddy simulations of 2D turbulence}

We consider two-dimensional flows expressed in the vorticity ($\omega$) and stream function ($\psi$) formulation in a doubly periodic square domain with length $L=2\pi$, 
\begin{subequations}\label{eq:NS}
  \begin{eqnarray}
    \frac{\partial \omega}{\partial t} + \mathcal{N}(\omega,\psi)&=&\frac{1}{\Rey}\nabla^2\omega - f -r\omega + \beta\frac{\partial \psi}{\partial x}, \label{eq:NS1}\\
    \nabla^2\psi &=& -\omega,
    \label{eq:NS2}
  \end{eqnarray}
\end{subequations}
where
$ \displaystyle
\mathcal{N}(\omega,\psi)=
\left(\nicefrac{\partial \psi}{\partial y}\right) \left(\nicefrac{\partial \omega}{\partial x}\right) - \left(\nicefrac{\partial \psi}{\partial x}\right) \left(\nicefrac{\partial \omega}{\partial y}\right),
$
is the nonlinear advection term,
and
$
f(x,y) = \kappa_f[\cos{(\kappa_fx)} + \cos{(\kappa_fy)}]
$
is a deterministic forcing~\cite{chandler2013invariant,kochkov2021machine}. To obtain target statistics, we solve Eq.~\eqref{eq:NS} numerically. A Fourier-Fourier pseudo-spectral solver is used to solve these equations, see e.g.~\cite{jakhar2023learning, guan2022stable,subel2022explaining}. The computational grid has uniform spacing $\Delta_\mathrm{DNS} = L/N_\mathrm{DNS}$, where $N_\mathrm{DNS}$ is the number of grid points in each direction, as shown in \cref{tab:cases}. The time-stepping size is $\Delta t_\mathrm{DNS}= 5\times10^{-5}$ dimensionless time unit for cases 1, 2, 3, 4 and $\Delta t_\mathrm{DNS}= 1\times10^{-5}$ dimensionless time unit for cases 5. Once the flow reaches statistical equilibrium after a long-term spin-up, we uniformly sample $5$ snapshots that are $10000\Delta t_\mathrm{DNS}$ apart to obtain decorrelated data to generate the reference enstrophy spectra, following~\cite{Guan_PhysicaD_2023,guan2024online}.

We perform \ac{FDNS} by applying a sharp Fourier cut-off filter (cut-off wavenumber $k_c=\pi/\Delta_{\text{\ac{FDNS}}}$) to each variable in \cref{eq:NS}, denoted by $\overline{\cdot}$, along each of $x$ and $y$ directions~\cite{pope2001turbulent,sagaut2006large}:
\begin{subequations}\label{eq:FNS}
  \begin{eqnarray}
    \frac{\partial \overline{\omega}}{\partial t} + \mathcal{N}(\overline{\omega},\overline{\psi})&=&\frac{1}{\Rey}\nabla^2\overline{\omega}-\overline{f}-r\overline{\omega}+\beta\frac{\partial \overline{\psi}}{\partial x}+\underbrace{\mathcal{N}(\overline{\omega},\overline{\psi}) - \overline{\mathcal{N}({\omega},{\psi})}}_{\Pi=\nabla\times(\nabla \cdot \tau^{\text{SGS}})}\label{eq:FNS1},\\
    \nabla^2\overline{\psi} &=& -\overline{\omega}\label{eq:FNS2}.
  \end{eqnarray}
\end{subequations}
The \ac{LES} equation is the same as Eq.~\eqref{eq:FNS} except for the unclosed SGS term $\Pi$. 
The \ac{LES} is solved on a coarse resolution $\Delta_\text{LES}=\Delta_\text{FDNS}$ with $\Pi$ modeled as a map connecting the resolved flow variables to the closure term.
The \ac{LES} is solved with the same numerical method as the \ac{DNS}, with grid spacing $\Delta_\mathrm{LES} = L/N_\mathrm{LES}$ and time step $\Delta t_\text{LES}=10\Delta t_\text{DNS}$.

\subsection{SMARL setup}

$N$ agents are distributed uniformly on the grid of  the simulation domain.
Agents can sense states $s^{(i)}(t)$ and perform actions according to a shared policy, $a^{(i)}(t) \sim \pi(\cdot | s^{(i)}(t))$. 
In this work, we use global states only, meaning that $s^{(i)}(t)=s'(t)$ for any agent $i$. 
In the appendix, we show that adding local states to the input of the policy does not improve the results.
The choice of using the same policy, $\pi(\cdot | s^{(i)}(t))$, for every agent is motivated by the translation invariance of the Navier-Stokes equations. The selection of states, actions, and rewards is critical to the performance of SMARL as detailed below.

\paragraph{States.} Each agent observes the enstrophy spectrum of the \ac{LES}, $s'(t)=\hat{Z}_\text{LES}(k,t)$.
We found  that adding the five Galilean-invariant quantities derived from filtered velocity gradients and Hessians~\cite{Ling_JFM_2016,novati2021automating} did not improve the performance of the closure model (\Cref{se:sobol}).

\paragraph{Action.} The agents control local coefficients $c_l(x,y,t)$ in traditional eddy-viscosity closures: Leith model~\cite{Leith_PhysicaD_1996}.
These coefficients are evaluated at the agents' locations and then interpolated onto the \ac{LES} grid.
The interpolated coefficients are then used to compute the subgrid stress tensor ${\tau}^{\text{SGS}}=-2\nu_e\bar{\mathcal{S}}$, which enters the \ac{LES} solver.
Actions are evaluated every 10 time steps of the LES solver, and local coefficients are kept constant between two consecutive actions.

\paragraph{Reward.} The objective of the agents is to learn a closure that matches the \ac{DNS} enstrophy spectrum.
Thus, we have chosen the instantaneous reward to be large when the enstrophy spectrum is close to the reference,
\begin{equation}
  r(t) = \frac{1}{\|\log \hat{Z}_\text{DNS}(k) - \log \hat{Z}_{\text{LES}}(k,t)\|^2_2}, \label{eq:reward}
\end{equation}
where $\hat{Z}_\text{DNS}$ is the spectrum obtained from \ac{DNS} simulations.
This reference spectrum is computed from short \ac{DNS} trajectories, which are known to be insufficient for purely offline training~\cite{Guan_PhysicaD_2023}.
Furthermore, these \ac{DNS} trajectories are much shorter than what would be required to capture the rare events that we report in the results section.

\subsection{Training algorithm: V-RACER}

The training of the policy is performed with the V-RACER algorithm with Remember and Forget experience replay (ReFER)~\cite{novati2019refer}, adapted to multi-agent settings as described in Weber et.~al.~\cite{weber2022remember} and implemented in the software Korali~\cite{Martin_CMAME_2022}.
Here, we describe the main components of the training algorithm.

At each time step $t$, agents observe the state $s'(t)$, perform an action sampled from the behavior policy $a(t) \sim \mu_t(\cdot, s'(t))$, and transition to a new state $s'(t+\Delta t)$ following the dynamics of the model (e.g., LES solver with \ac{SMARL}-based closure here), with reward $r(t+\Delta t)$.
Experiences $\{s'(t), a(t), s'(t+\Delta t), r(t), \mu_t\}$ are stored in a replay buffer.
In V-RACER, the policy $\pi^\theta(\cdot, s)$ is represented as a Gaussian with mean $m^\theta(s)$ and covariance $\Sigma^\theta(s)$, which, in addition to the value $V^\theta(s)$, are the output of a single neural network with weights $\theta$ and input $s$.

The weights $\theta$ are updated to minimize two losses, corresponding to the off-policy objective and to match the value function with an on-policy value estimate.
We use the Adam gradient-based optimization method to minimize these objectives.
The off-policy objective is used to update the policy weights:
\begin{equation}
  \mathcal L^\text{off-PG}(\theta) = - \mathbb{E}_{s_k \sim B, a_k\sim \mu_k(s_k)} \left[ \rho_k \left( \hat{Q}_t^\text{ret} - V^\theta(s_k) \right) \right],
\end{equation}
where $B(s) \propto \sum\limits_{k=t-N}^t P(s_k=s| s_0, a_k \sim\mu_k)$ is the probability of sampling the state $s$ from the replay memory containing the past $N$ experiences of the agent acting with policy $\mu_k$.
Furthermore, the importance weight is defined as $\rho_k = \pi^\theta(s_k|a_k) / \mu_k(s_k|a_k)$, and the on-policy returns are estimated as
\begin{equation}
  \hat{Q}_{t-1}^\text{ret} = r_t + \gamma V^\theta(s_t) | \gamma \bar{\rho}_t \left[ \hat{Q}_{t}^\text{ret} - V^\theta(s_t) \right],
\end{equation}
with $\bar{\rho}_t = \min \{1, \rho_t\}$.
The on-policy state-value is estimated from $V^\theta$ using the recursive formula~\cite{novati2019refer}
\begin{equation}
  \hat{V}_t^\text{tbc} = V^\theta(s_t) + \bar{\rho}_t \left[r_{t+1} + \gamma \hat{V}_{t+1}^\text{tbc} - V^\theta(s_t) \right].
\end{equation}
This estimate is part of the loss used to train the value function:
\begin{equation}
  \mathcal{L}^\text{tbc}(\theta) = \frac 1 2 \mathbb{E}_{s_k \sim B, a_k\sim \mu_k(s_k)} \left[ V^\theta(s_t) - \hat{V}_t^\text{tbc}\right]^2.
\end{equation}
The gradients of these losses are updated according to the ReF-ER procedure~\cite{novati2019refer}.
This consists of clipping gradients to zero when the importance weight is outside of the interval $[1/c_\text{max}, c_\text{max}]$, where $c_\text{max}$ is a positive constant.
Furthermore, the gradients are biased towards the direction of past behaviors,
\begin{equation}
 \mathbf{g}_t^\text{ReF-ER}(\theta) = \beta \mathbf{g}_t(\theta) - (1-\beta) \nabla D_\mathrm{KL}   \left[\mu_t(\cdot|s_t) || \pi^\theta(\cdot|s_t)\right],
\end{equation}
where $\beta>0$ is a hyper-parameter and $D_\mathrm{KL}$ represents the KL-divergence between two distributions.

\subsection{State-action map sensitivity analysis}

To determine the effects of the \ac{SMARL} \ac{DNN} input ($\hat{Z}(k,t)$) at each wavenumber $k$ to the NN output ($c_l(t)=f_\text{DNN}(\hat{Z}(k,t))$), we use the Sobol index which decomposes the variance of the output into contributions from: (1) each individual input, and (2) interaction among inputs. The variance of $c_l$ can be decomposed as:
\begin{eqnarray}
    Var(c_l) = \sum_m\xi_m+\sum_{m<n}\xi_{mn}+\cdots+\xi_{1,2,3,...,k_{c}},
\end{eqnarray}
where here $\xi_m=Var\big(\mathbb{E}(c_l|\hat{Z}(m))\big)$ are the partial variances associated with one component of the decomposition. The first-order Sobol index is defined as
\begin{eqnarray}\label{eq:sobol_num_1}
    S_j = \frac{\xi_j}{Var(c_l)},
\end{eqnarray}
which is the fraction of output variance explained by varying $\hat{Z}(k=k_j)$ alone. The total index:
\begin{eqnarray}\label{eq:sobol_num_total}
    S_{T_j} = 1- \frac{\xi_{\sim j}}{Var(c_l)},
\end{eqnarray}
which the fraction of output variance explained by varying $\hat{Z}(k_j)$ and all interactions involving it. $\xi_{\sim j}$ is the variance of $c_l$ while holding $\hat{Z}(k_j)$ fixed.

\subsection{Experiments and test cases}
\label{sec:expeiments}

We have developed closures for 5 different forced 2D turbulent flows (2 with the $\beta$-plane effects; details in \cref{tab:cases}).
These cases are commonly used to evaluate the \ac{SGS} closures of geophysical turbulence~\cite{frezat2021physical, Guan_PhysicaD_2023, srinivasan2024turbulence} and exhibit distinct behaviors and dynamics, as seen in snapshots of vorticity fields, $\omega$, in \cref{fig:overview}.
Training is performed with the objective of achieving an \ac{LES} enstrophy spectrum close to the target DNS spectrum.
We also compare the kinetic energy spectra in \cref{sec:aposteriori:kinetic:spectra}.
Additionally, to test the predictability of \ac{SMARL}-based LES models for extreme events, we compare the PDFs of the resolved vorticity, $\mathcal{P}(\omega)$.
The tails of these PDFs represent rare, extreme events, i.e., significantly large $\omega$ with a small probability of occurrence.
Note that these vortices resemble the weather system's high- and low-pressure anomalies, which can cause various extreme weather events~\cite{woollings2018blocking}.
LES with $16$ to $16,384\times$ coarser spatial resolution and $10\times$ larger time steps coupled to the learned closures are then tested with the \ac{SMARL}-based closures, and their statistics are compared with those of the DNS and LES with classical dynamic Smagorinsky and Leith closures.
As summarized in \cref{fig:aposteriori:testing}, the tails of the vorticity PDFs clearly show the advantage of the \ac{SMARL}-based closures, suggesting that these closures have the right amount of diffusion and backscattering (anti-diffusion).
The classical closures are too diffusive and underestimate extreme events.
The energy spectra also show the better ability of LES with \ac{SMARL}-based closures in capturing the energy across the scales.

\label{se:cases}

\def\deltatrl{\ensuremath{\Delta t_\text{RL}}}
\def\deltatdns{\ensuremath{\Delta t_\text{LES}}}
\begin{table}[!tbh]
  \caption{The test cases and hyper-parameters in training. $\sigma_\omega$ is standard deviation of vorticity.}
  \label{tab:cases}
  \centering
  \begin{tabular}{c|cccc|ccccc}
    \toprule
    Case & $\Rey$ & $\beta$ & $\kappa_f$ & $\sigma_\omega$ & $N_\text{DNS}$&$N_\text{LES}$& $\deltatrl/\deltatdns$&  Training horizon & Testing horizon\\
    \midrule
    1 & $20\times10^3$ & $0$  & $4$&$5.51$&$1024$&$32$&$10$&$10^3\deltatrl$& $10^6\deltatrl$\\
    2 & $20\times10^3$ & $20$ & $4$ &$10.75$ &$1024$&$32$&$10$&$10^3\deltatrl$& $10^6\deltatrl$\\
    3 & $20\times10^3$ & $50$ & $10$ &$13.14$&$1024$&$128$&$10$&$10^3\deltatrl$& $10^6\deltatrl$\\
    4 & $20\times10^3$ & $0$  & $25$ &$13.47$&$1024$&$256$&$10$&$10^3\deltatrl$& $10^6\deltatrl$\\
    5 & $300\times10^3$ & $0$ & $4$  &$6.42$&$4096$&$32$&$10$&$10^3\deltatrl$& $10^6\deltatrl$\\
    \bottomrule
  \end{tabular}
\end{table}

\bibliographystyle{unsrt}
\bibliography{references.bib, refPH, refGM, refs_aires}

\appendix

\section{A-posteriori analysis in turbulent kinetic energy spectra}
\label{sec:aposteriori:kinetic:spectra}

In this section, we compare the turbulent kinetic energy spectra, $\hat{E}(k)$, of the LES with \ac{SMARL}-based (RL-Smag and RL-Leith) and traditional dynamic Smag and Leith (DSmag and DLeith) models with the ones obtained from DNS. The \ac{SMARL}-based models match the DNS/FDNS spectra over the whole spectrum $k\in[0,k_c]$ across all 4 cases, outperforming both the DSmag and DLeith models.

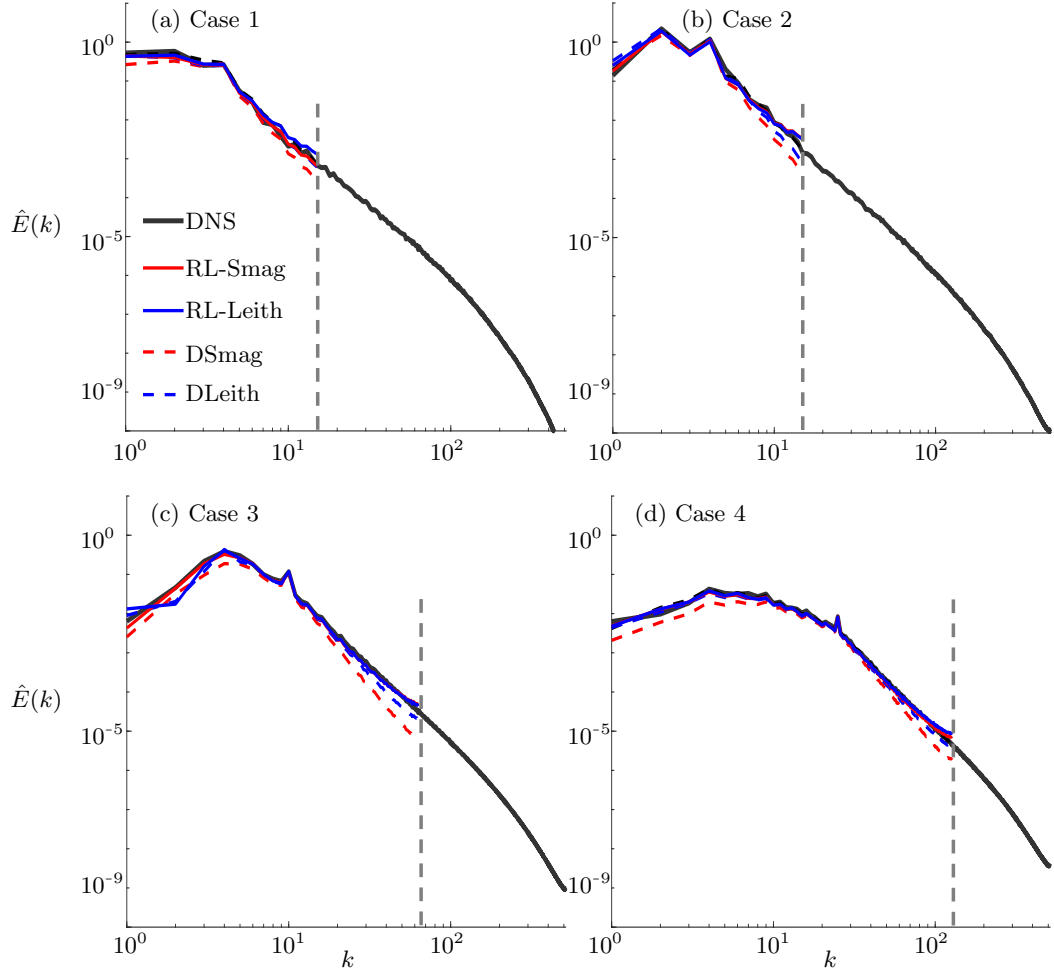
\begin{figure}
 \centering
 \begin{overpic}
[width=0.75\linewidth,height=0.75\linewidth]{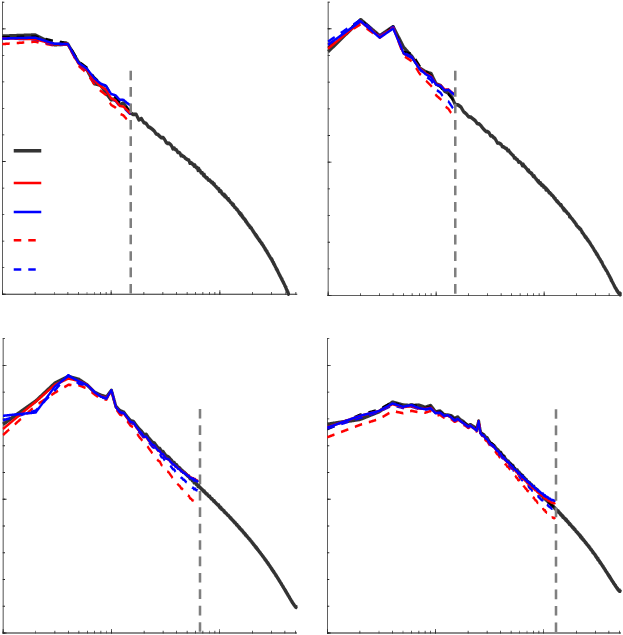}

\put(3,97){\small (a) Case 1}
\put(60,97){\small (b) Case 2}
\put(3,44){\small (c) Case 3}
\put(55,44){\small (d) Case 4}

\put(6.8,75.3){\small DNS}
\put(6.8,70.3){\small RL-Smag}
\put(6.8,65.8){\small RL-Leith}
\put(6.8,61.0){\small DSmag}
\put(6.8,56.8){\small DLeith}

\put(-1,50.5){\small $10^0$}
\put(16,50.5){\small $10^{1}$}
\put(33,50.5){\small $10^{2}$}

\put(-4.5,57){\small $10^{\shortminus 9}$}
\put(-4.5,73){\small $10^{\shortminus 5}$}
\put(-4.5,94){\small $10^{0}$}

\put(50,50.5){\small $10^0$}
\put(68,50.5){\small $10^{1}$}
\put(85,50.5){\small $10^{2}$}

\put(47.8,57){\small $10^{\shortminus 9}$}
\put(47.8,73){\small $10^{\shortminus 5}$}
\put(47.8,94){\small $10^{0}$}

\put(-1,-2.5){\small $10^0$}
\put(16,-2.5){\small $10^{1}$}
\put(33,-2.5){\small $10^{2}$}

\put(-4.5,4){\small $10^{\shortminus 9}$}
\put(-4.5,20){\small $10^{\shortminus 5}$}
\put(-4.5,41){\small $10^{0}$}

\put(50,-2.5){\small $10^0$}
\put(68,-2.5){\small $10^{1}$}
\put(85,-2.5){\small $10^{2}$}

\put(47.8,4){\small $10^{\shortminus 9}$}
\put(47.8,20){\small $10^{\shortminus 5}$}
\put(47.8,41){\small $10^{0}$}

\put(23.5,-4){\small $k$}
\put(75.3,-4){\small $k$}

\put(-12,24){\small $\hat{E}(k)$}
\put(-12,75){\small $\hat{E}(k)$}

\end{overpic}

\vspace*{5mm}

 \caption{
   {\it A-posteriori} testing results in terms of turbulent kinetic energy spectra, $\hat{E}(k)$. The gray dash lines mark the cut-off wavenumber, $k_c$. The FDNS spectra is the same as the DNS spectra up to $k_c$.}
 \label{fig:aposteriori:testing_local}
\end{figure}

\section{Sensitivity analysis of the policy}
\label{se:sobol}

In this section, we investigate which states are responsible for the most change in the output of the policy.
To do so, we train the RL agents in cases 1, 2, 3, and 4 (see \cref{se:cases}) with both the Smagorinsky and Leith models.
The input states contain the five nonzero local invariants of the velocity gradient and velocity Hessian, followed by the logarithm of the enstrophy spectrum.
We then compute the first-order Sobol indices of the policy for each input state, using the SALib package~\cite{iwanaga2022toward}.



\begin{figure}
 \centering
 \begin{overpic}
[width=0.75\linewidth,height=0.75\linewidth]{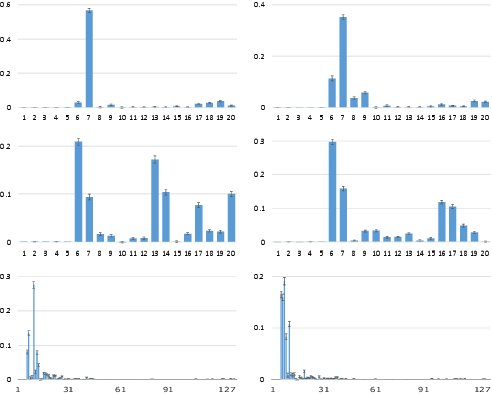}

\put(18,102){RL-Leith}
\put(70,102){RL-Smag}

\end{overpic}
\vspace*{5mm}

 \caption{
   First-order Sobol indices of the policies trained with RL-Leith and RL-Smag for all (local + global) states in Cases 1, 2, and 4 (from top to bottom).}
 \label{fig:SA:CL_CS}
\end{figure}

\Cref{fig:SA:CL_CS} shows the first-order Sobol indices for all three cases with the RL-Leith and RL-Smagorinsky models, repectively.
We observe that in all cases, the first five states, corresponding to the local invariants of the velocity gradient (2 local invariants in 2D turbulence) and velocity Hessian (3 local invariants in 2D turbulence) ~\cite{novati2021automating}, have a nearly zero first-order Sobol index.
This means that the policy does not depend on these states, and we have thus chosen to remove them from the input of the policies in this study.
We also note that in most cases, input states that correspond to low frequencies of the enstrophy spectrum have a dominant first-order Sobol index, indicating that high frequencies do not play a major role in the control of the coefficients $c_s$ and $c_l$.

\begin{figure}
 \centering
 \begin{overpic}
[width=0.75\linewidth,height=0.75\linewidth]{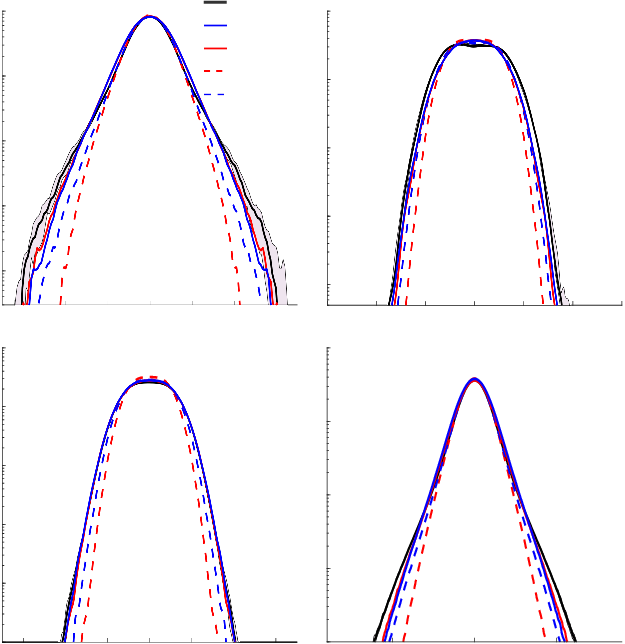}

\put(3,97){\small (a) Case 1}
\put(55,97){\small (b) Case 2}
\put(3,44){\small (c) Case 3}
\put(55,44){\small (d) Case 4}

\put(36.5,98.8){\small DNS}
\put(36.5,95.3){\small RL-Leith}
\put(36.5,91.5){\small RL-Smag}
\put(36.5,88.0){\small DSmag}
\put(36.5,84.5){\small DLeith}

\put(1,50.5){\small $\shortminus 6$}
\put(9,50.5){\small $\shortminus 4$}
\put(16,50.5){\small $\shortminus 2$}
\put(23.5,50.5){\small $0$}
\put(30,50.5){\small $2$}
\put(37,50.5){\small $4$}
\put(44,50.5){\small $6$}

\put(-4.5,57){\small $10^{\shortminus 5}$}
\put(-4.5,67){\small $10^{\shortminus 4}$}
\put(-4.5,77){\small $10^{\shortminus 3}$}
\put(-4.5,87){\small $10^{\shortminus 2}$}
\put(-4.5,97){\small $10^{\shortminus 1}$}

\put(59,50.5){\small $\shortminus 4$}
\put(66.5,50.5){\small $\shortminus 2$}
\put(75.3,50.5){\small $0$}
\put(83,50.5){\small $2$}
\put(91,50.5){\small $4$}

\put(47.2,55){\small $10^{\shortminus 5}$}
\put(47.2,65.5){\small $10^{\shortminus 4}$}
\put(47.2,76){\small $10^{\shortminus 3}$}
\put(47.2,87){\small $10^{\shortminus 2}$}
\put(47.2,97){\small $10^{\shortminus 1}$}

\put(1.7,-2){\small $\shortminus 6$}
\put(9,-2){\small $\shortminus 4$}
\put(16,-2){\small $\shortminus 2$}
\put(23.5,-2){\small $0$}
\put(30,-2){\small $2$}
\put(37,-2){\small $4$}
\put(43.7,-2){\small $6$}

\put(-4.5,8.5){\small $10^{\shortminus 5}$}
\put(-4.5,17.6){\small $10^{\shortminus 4}$}
\put(-4.5,26.7){\small $10^{\shortminus 3}$}
\put(-4.5,36){\small $10^{\shortminus 2}$}
\put(-4.5,45){\small $10^{\shortminus 1}$}

\put(61,-2){\small $\shortminus 5$}
\put(75.3,-2){\small $0$}
\put(88.5,-2){\small $5$}

\put(47.2,11){\small $10^{\shortminus 4}$}
\put(47.2,22){\small $10^{\shortminus 3}$}
\put(47.2,33){\small $10^{\shortminus 2}$}
\put(47.2,44){\small $10^{\shortminus 1}$}

\put(23.5,-4){\small $\omega/\sigma_\omega$}
\put(75.3,-4){\small $\omega/\sigma_\omega$}
\put(23.5,48.5){\small $\omega/\sigma_\omega$}
\put(75.3,48.5){\small $\omega/\sigma_\omega$}

\put(-12,25){\small $\mathcal{P}(\omega)$}
\put(-12,75){\small $\mathcal{P}(\omega)$}

\end{overpic}

\vspace*{5mm}

 \caption{
   {\it A-posteriori} testing results in terms of probability density function of vorticity ($\mathcal{P}(\omega)$). The $x-$axis is normalized by the standard deviation of vorticity ($\sigma_\omega$) computed from DNS of each case. The shading area represents the uncertainty (25-75 quartiles) of the DNS PDF. Here, the RL models take both local states (5 invariants of velocity gradient and velocity Hessian tensors) and global states (logarithm of enstrophy spectrum) as inputs.}
 \label{sfig:aposteriori:testing_local}
\end{figure}
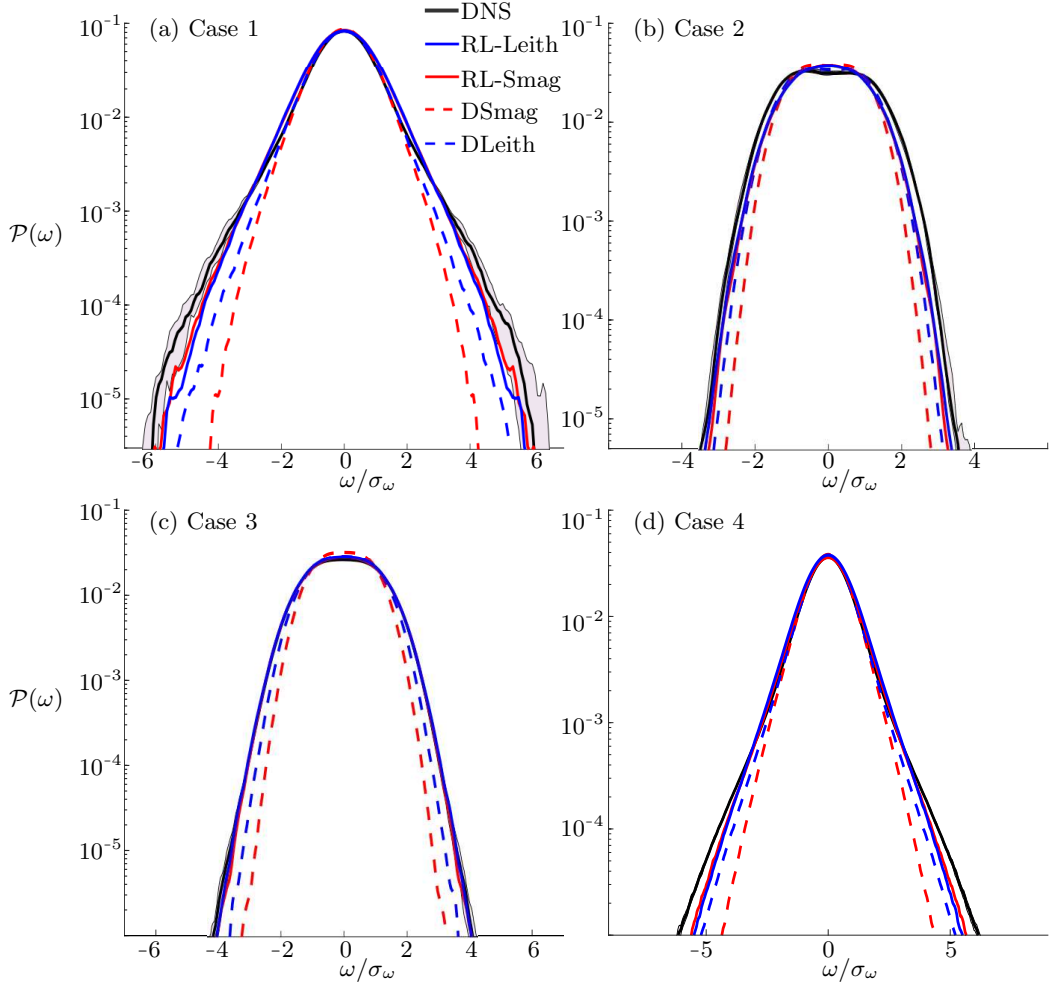

The effects of local states on the {\it a-posteriori} results are shown in \Cref{sfig:aposteriori:testing_local}. In terms of predicting the probability density function of vorticity, $\mathcal{P}(\omega)$, the RL models with local states perform similarly to those with global states only (see \cref{fig:aposteriori:testing}). Another note is that Smag and Leith perform similarly with RL optimized coefficients, which is reasonable due to the equivalence of both models in 2D once their coefficients are optimized~\cite{guan2024online,guan2025semi}.

\end{document}